\documentclass[%
 aip,
 jmp,%
 amsmath,amssymb,
preprint,%
]{revtex4-1}

\draft % marks overfull lines with a black rule on the right

\usepackage{graphicx}% Include figure files
\usepackage{dcolumn}% Align table columns on decimal point
\usepackage{bm}% bold math
\usepackage{amsmath}
\usepackage{amsthm}
\newcommand{\dif}{\mathrm{d}}
\newcommand{\ue}{\mathrm{e}}
\newcommand{\ui}{\mathrm{i}}
\DeclareMathAlphabet{\mathsfsl}{OT1}{cmss}{m}{sl}

\begin{document}
\newtheorem{definition}{Definition}[section]
\newtheorem{theorem}{Theorem}[section]
\newtheorem{lemma}{Lemma}[section]
\newtheorem{proposition}{Proposition}[section]

%\preprint{}

\title{Quantization of Damping Particle Based On New Variational Principles}


\author{Tianshu Luo}
\email[]{ltsmechanic@zju.edu.cn}
\author{Yimu Guo}
\email[]{guoyimu@zju.edu.cn}
\affiliation{Institute of Solid Mechanics, Department of Applied Mechanics, Zhejiang University,\\
Hangzhou, Zhejiang, 310027,  P.R.China\\
}

\date{\today}

\begin{abstract}

In this paper a new approach is proposed to quantize mechanical systems whose equations of motion can not be put into Hamiltonian form.
This approach is based on a new type of variational principle, which is adopted to a describe a relation: a damping particle may shares a 
common phase curve with a free particle, whose Lagrangian in the new variational principle can be considered as a Lagrangian density in 
phase space. According to Feynman's theory, the least action principle is adopted to modify the Feynman's path integral formula, where 
Lagrangian is replaced by Lagrangian density. In the case of conservative systems, the modification reduces to standard Feynman's propagator 
formula. As an example a particle with friction is analyzed in detail.
\end{abstract}

\pacs{01.70.+w, 02.40.Yy, 03.65.Ca, 45.20.-d}
\keywords{Quantization, Damping, Dissipative,Path-integral, Propagator, Variational principle}
%\keywords{}

\maketitle

\section{Introduction}
Quantum mechanics is more rigorous than the classical one, which is the best elaborated and understood part of physics. Classical mechanics can be thought of 
as the base of quantum mechanics. Lagrangian Mechanics and Hamiltonian mechanics are geometrical description of classical mechanics. 
In physics, quantization is the process of explaining a classical understanding of physical phenomena in terms of a newer understanding known as 
quantum mechanics. This is a generalization of the procedure for building quantum mechanics from classical mechanics. If a physical phenomenon posses Lagrangian 
or Hamiltonian description, the quantization procedure can be easy performed. But for damping classical physical phenomena, it is complicated to perform a 
quantization procedure from the classical description to the quantum one. Readers can read the complete history of the important ideas in this field in the 
literature\cite{2005RvMaP..17..391A}. Recent works on quantization of damped systems see papers\cite{1986AmJPh..54..273H,Das2005,
springerlink:10.1007/s10773-006-9064-9,Chruscinski2006854,Blasone2004354,PhysRevA.68.014101,chandrasekar:032701,Dito2006309}.

The works of Kochan\cite{Kochan2010219}\cite{PhysRevA.81.022112} \cite{2009IJMPA..24.5319K}\cite{2007hep.th....3073K}attracts our attention, 
because there is close link between the works and the classical mechanics. Kochan\cite{Kochan2010219} utilized the tools of contact geometry
 to give a new picture of classical mechanics, and then defined a new variational principle of least action\cite{2007hep.th....3073K}. Based on 
the classical picture and the variational principle, a quantum propagator\cite{2007hep.th....3073K} for damping systems was derived.

The ideas of Kochan motivated us to turn our interest to quantum mechanics, because we have found a relation between a damping classical mechanical 
system and conservative classical ones, which can be described as a variational principle\cite{2009arXiv0906.3062L}. This picture of classical mechanics 
 is illustrated by a simple example in Sec. II. Therefore, we attempt to utilize our variational principle\cite{2009arXiv0906.3062L} to construct a new quantum 
propagator. Our approach to define a new propagator is explained in Sec. III. In Sec. IV a simple example, where the quantization of a particle 
with friction $\kappa \dot{x}$ is performed, is reported.  In this example, comparison with several other approach is made.

\section{Review On a New Variational Principle}
\subsection{A Proposition}
To beging, it is necessary to review the new variational principle\cite{2009arXiv0906.3062L}, which is adopted to describe a relation between damping classical 
mechanical system and some conservative ones. The relation is represented as a proposition:
\begin{proposition}
For any non-conservative classical mechanical system and arbitrary initial condition, there exists a conservative system; both systems sharing 
one and only one common phase curve; and the value of the Hamiltonian of the conservative system is equal to the sum of the total energy 
of the non-conservative system on the aforementioned phase curve and a constant depending on the initial condition.
\label{pro:1}
\end{proposition}
This proposition can be demonstrated by a simple example.
Consider a special one-dimensional simple mechanical system
\begin{equation}
 \ddot{x}+\kappa\dot{x}=0,
\label{eq:simp_1d}
\end{equation}
where $c$ is a constant. The exact solution of the equation above is
\begin{equation}
 x=\Lambda+\eta \ue^{-\kappa t},
\label{eq:sol_1d}
\end{equation}
where $\Lambda,\eta$ are constants. Differentiation gives the velocity:
\begin{equation}
 \dot{x}=-\kappa \eta \ue^{-\kappa t}.
\label{eq:sol_1dv}
\end{equation}
From the initial condition $x_0,\dot{x}_0$, we find $\Lambda=x_0+\dot{x}_0/\kappa,\ \ \eta=-\dot{x}_0/\kappa$. 
Inverting Eq. (\ref{eq:sol_1d}) yields
\begin{equation}
t=-\frac{1}{\kappa}\ln\frac{x-\Lambda}{\eta}
\label{eq:tfunc}
\end{equation}
and by substituting into Eq. (\ref{eq:sol_1dv}), such we have
\begin{equation}
 \dot{x}=-\kappa(x-\Lambda)
\label{eq:dx}
\end{equation}
The dissipative force $F$ in the dissipative system (\ref{eq:simp_1d}) is
\begin{equation}
 F=\kappa\dot{x}.
\label{eq:F}
\end{equation}
Substituting Eq. (\ref{eq:dx}) into Eq. (\ref{eq:F}), the conservative force $\mathcal{F}$ is expressed as
\begin{equation}
  \mathcal{F}=-\kappa^2(x-\Lambda);
\label{eq:mF}
\end{equation}
Clearly, the conservative force $\mathcal{F}$ depends on the initial condition of the dissipative system (\ref{eq:simp_1d}), in other words,  
an initial condition determines a conservative force. Consequently, a new conservative system yields
\begin{equation}
 \ddot{x}+\mathcal{F}=0\rightarrow \ddot{x}-\kappa^2(x-\Lambda)=0.
\label{eq:1d_eq_consys}
\end{equation}
The stiffness coefficient in this equation must be negative. One can readily verify that the particular solution (\ref{eq:sol_1d}) 
of the dissipative system can satisfy the conservative one (\ref{eq:1d_eq_consys}). This point agrees with Proposition (\ref{pro:1}).

The Lagrangian of the system (\ref{eq:1d_eq_consys}) is
\begin{equation}
 \hat{L}=\frac{1}{2}\dot{x}^2+\frac{1}{2}\kappa^2x^2-\kappa^2\Lambda x
\label{eq:1d_eq_Lag}
\end{equation}
\subsection{Infinite-dimensional Variational Principle}
 The ideas of literatures\cite{RickSalmon1988,RevModPhys.70.467,Morrison1980} have motivated us to regard the mechanical 
system (\ref{eq:simp_1d}) as a special fluid, which is a collection of fluid particles in phase space. Therefore, 
first let the label of a particle in the phase space be
\begin{equation}
 \bm{a}=(x_0,\dot{x}_0);
\label{eq:lag_vrb}
\end{equation}
the coordinate of a particle in the configuration space
\begin{equation}
 \bm{q}=\bm{q}(\bm{a},t)=\left\lbrace x(\bm{a},t),\dot{x}(\bm{a},t)\right\rbrace ;
\end{equation}
$\rho_o=1$.
One can consider $\hat{L}$ in Eq. (\ref{eq:1d_eq_Lag}) as a Lagrangian 
density of the system (\ref{eq:simp_1d})
\begin{equation}
\mathcal{L}=\hat{L}=\frac{1}{2}\dot{x}^2+\frac{1}{2}\kappa^2x^2-\kappa^2\Lambda x
\label{eq:Lag_denst}
\end{equation}
Thus the Lagrangian functional of Eq. (\ref{eq:simp_1d}) can be presented as the following:
\begin{equation}
L[x,\dot{x}]=\int_D \mathcal{L}\dif^{2} \bm{a}=\int_D\left[\frac{1}{2}\dot{x}^2+\frac{1}{2}\kappa^2x^2-\kappa^2\Lambda x \right]\dif^{2} \bm{a},
\label{eq:nh2}
\end{equation}
where $\dif^{2}=\dif x_0\dif \dot{x}_0$.  
Thus the action functional can be presented as follows:
\begin{equation}
 S[\bm{q}]=\int^{t1}_{t0}L[x,\dot{x}]\dif t
=\int^{t1}_{t0}\dif t\int_D \left[\frac{1}{2}\dot{x}^2+\frac{1}{2}\kappa^2x^2-\kappa^2\Lambda x \right]\dif^{2}\bm{a}
\label{eq:nh3}
\end{equation}
According to Hamiltonian theorem, we have the functional derivative $\delta S/\delta \bm{q}(a,t)=0$,
\begin{eqnarray}
 \frac{\delta S}{\delta \bm{q}(\bm{a},t)}=0 &\Longrightarrow& \frac{\delta L}{\delta \bm{q}(\bm{a},t)}
-\frac{\dif}{\dif t}\frac{\delta L}{\delta \dot{\bm{q}}(\bm{a},t)}=0\nonumber\\
&\Longrightarrow&\frac{\partial \mathcal{L}}{\partial \bm{q}(\bm{a},t)}
-\frac{\dif}{\dif t}\frac{\partial\mathcal{L}}{\partial \dot{\bm{q}}(\bm{a},t)}=-\ddot{x}(\bm{a},t)-\kappa^2(x-\Lambda)=0 .
\label{eq:nh4}
\end{eqnarray}
The equation above implies that subject to the initial condition $\bm{a}$, an associated conservative system exists, the control equation of which
is Eq. (\ref{eq:1d_eq_consys}), the phase curve of which coincides with that of the damping system (\ref{eq:simp_1d}). In the case of the classical least action principle, 
the stationary curves set is represented as an Euler-Lagrangian differential equation, which can be converted into a uniform conservative Newtonian equation. 
In the case of our least action principle, the stationary curves set is represented as an Euler-Lagrangian functional equation, which can be converted varied conservative 
Newtonian equation subject to varied initial condition $\bm{a}$.

Furthermore we can define an action density $\hat{S}$, such that Eq. (\ref{eq:nh3}) can be rewritten as
\begin{equation}
  S[\bm{q}]=\int_D \hat{S}\dif^{2}\bm{a}=\int_D \dif^2\bm{a}\int^{t1}_{t0} \mathcal{L} \dif t,
\end{equation}
where $\hat{S}=\int^{t1}_{t0} \mathcal{L} \dif t$. If Eq. (\ref{eq:simp_1d}) is conservative, the action density reduces back to the classical action, 
and the Lagrangian density back to a uniform classical Lagrangian, our least action principle back to the classical least action principle.

\section{Derivation of a Quantum Propagator}
According to Feynman's theory\cite{Feynman1965}, the probability amplitude of the transition of a system from the space-time configuration $(x_a,0)$ to another space-time 
configuration $(x_b,T)$ is
\begin{eqnarray}
 K(x_b,T;x_a,0)&=&\frac{1}{N}\int \left[\mathcal{D} \gamma\right]\exp\left\lbrace\frac{\ui}{\hbar}S\right\rbrace \nonumber\\
&=&\frac{1}{N}\int \left[\mathcal{D} \gamma\right]\exp\left\lbrace\frac{\ui}{\hbar}\int_\gamma L \dif t\right\rbrace 
\label{eq:Fm_propg}
\end{eqnarray}
Here the integral is taken over all pathes $\gamma(t)=(x,t)$ or all pathes in the phase space, satisfying the boundary conditions 
\begin{equation}
x(0)=x_a,\ \ x(T)=x_b.
\label{eq:bndcondt}
\end{equation}
Kochan employed his variational principle to replace $L \dif t$ in Feynman's kernel function with a Lepage two-form $\Omega$. Motivated by the Kochan's trick
\cite{2007hep.th....3073K}, encouraged by the sentence from Feynman's thesis\cite{Feynman_thesis} too: ``the central mathematical concept is the analogue of the action in 
classical mechanics. It is therefore applicable to mechanical systems whose equations of motion cannot be put into Hamiltonian form. It is only 
required that some sort of least action principle be available'', we propose a method to generalize the Feynman's probability amplitude:
\begin{eqnarray}
 K(x_b,T;x_a,0)&=&\frac{1}{N}\int \left[\mathcal{D} \gamma\right]\exp\left\lbrace\frac{\ui}{\hbar}\hat{S}\right\rbrace \nonumber\\
&=&\frac{1}{N}\int \left[\mathcal{D} \gamma\right]\exp\left\lbrace\frac{\ui}{\hbar}\int_\gamma \mathcal{L} \dif t\right\rbrace, 
\label{eq:G_Fm_propg}
\end{eqnarray}
where the action $S$ is replaced by the action density $\hat{S}$ and the Lagrangian $L$ replaced by the Lagrangian density $\mathcal{L}$, the classical 
least action principle is replaced by the variational principle in Sec. III.

In conservative case, the Lagrangian density $\mathcal{L}$ reduces back to standard Lagrangian, the propagator (\ref{eq:G_Fm_propg}) reduces back to 
standard Feynman's propagator (\ref{eq:Fm_propg})

\section{Example of Quantization}
Consider the quantization of the system (\ref{eq:simp_1d}) with the boundary condition (\ref{eq:bndcondt}). 
From the boundary condition can be derived
\begin{equation}
 \Lambda=x_a-\frac{x_b-x_a}{\ue^{-\kappa T}-1},\ \ \eta=\frac{x_b-x_a}{\ue^{-\kappa T}-1}
\end{equation}
 
We take Feynman's general method\cite{Feynman1965} to integrate Eq. (\ref{eq:G_Fm_propg}). First we construct a short-time propagator:
\begin{eqnarray}
 K(x_{j+1},x_j;\epsilon)&=&\sqrt{\frac{1}{2\pi i\hbar}}\exp\left\lbrace \frac{\ui\epsilon}{\hbar}\left[\frac{(x_{j+1}-x_j)^2}{2\epsilon^2}
+\frac{\kappa^2}{4}\left(x_{j+1}^2+x_j^2 \right)-\kappa^2 \Lambda(\frac{x_{j+1}+x_j}{2})\right]\right\rbrace \nonumber\\
&=&\sqrt{\frac{1}{2\pi \ui\hbar}}\exp\left\lbrace\frac{\ui}{\hbar}
\left[ a_0(x_{j+1}^2+x_j^2)-2b_0x_{j+1}x_j-R_0x_{j+1}-S_0x_j\right]\right\rbrace,
\label{eq:short_prpgt}
\end{eqnarray}
where
\begin{equation}
 a_0=\frac{1}{2\epsilon}(\frac{1}{\epsilon}+\frac{(\kappa\epsilon)^2}{2}),\ \ \ \,b_0=\frac{1}{2\epsilon},\ \ \ \ R_0=\frac{\kappa^2 \Lambda\epsilon}{2} \ \ \ \ S_0=\frac{\kappa^2 \Lambda\epsilon}{2}.
\end{equation}
Substituting the short-time propagator into 
\[
 K(x_b,T;x_a,0)=\prod_{k=1}^{N-1}\int_{-\infty}^{+\infty}\dif x_k\prod_{j=0}^{N-1}K(x_{j+1},x_j;\epsilon),
\]
we have
\begin{eqnarray}
K(x_b,T;x_a,0)&=&\lim_{\substack{N \rightarrow \infty\\ \epsilon\rightarrow 0}}
\left(\frac{1}{2\pi \ui\hbar\epsilon}\right)^{N/2}\left[\prod_{j=1}^{N-1} \int \dif x_j\right]\nonumber\\
&&\times\exp\left\lbrace\frac{\ui}{\hbar}\sum_{j=0}^{N-1}\left[ a_0(x_{j+1}^2+x_j^2)-2b_0x_{j+1}x_j-R_0x_{j+1}-S_0x_j\right]\right\rbrace \nonumber\\
&=&\lim_{N\rightarrow \infty}A_1\left[\prod_{j=1}^{N-1} \int \dif x_j\right]\exp\left\lbrace\frac{\ui}{\hbar}\phi_1 \right\rbrace \nonumber\\
&=&\lim_{N\rightarrow \infty}A_2\left[\prod_{j=2}^{N-1} \int \dif x_j\right]\exp\left\lbrace\frac{\ui}{\hbar}\phi_2 \right\rbrace \nonumber\\
&=&\lim_{N\rightarrow \infty}A_k\left[\prod_{j=k}^{N-1} \int \dif x_j\right]\exp\left\lbrace\frac{\ui}{\hbar}\phi_k \right\rbrace \nonumber\\
&=&\lim_{N\rightarrow \infty}A_N\exp\left\lbrace\frac{\ui}{\hbar}\phi_N \right\rbrace.
\label{eq:dp_prpgt_1}
\end{eqnarray}
Here
\[
 A_1=\left(\frac{1}{2\pi \ui\hbar\epsilon}\right)^{N/2}
\]
\[
 \phi_1=\sum_{j=2}^{N-1}\left[a_0(x_{j+1}^2+x_j^2)-2b_0x_{j+1}x_j\right]+\alpha_1
\]
\begin{eqnarray*}
 \alpha_1&=&a_0(x_2^2+x_1^2)-2b_0x_2x_1-R_0x_2-S_0x_1+a_0(x_1^2+x_0^2)-2b_0x_1x_0-R_0x_1-S_0x_0 \nonumber\\
&=&a_0(x_2^2+x_0^2)-R_0x_2-S_0x_0+2a_0\left[ x_1-\frac{b_0(x_2+x_0)^2+(S_0+R_0)/2}{2a_0}\right]^2-\\
&&\frac{\left[b_0(x_2+x_0)^2+(S_0+R_0)/2\right]^2}{2a_0}\nonumber\\
&=&2a_0\left[ x_1-\frac{b_0(x_2+x_0)^2+(S_0+R_0)/2}{2a_0}\right]^2+a_1(x_2^2+x_0^2)-2b_1x_2x_0-R_1x_2-S_1x_0-\Omega_1\\
\end{eqnarray*}
where 
\begin{eqnarray}
 &&a_1=a_0-\frac{b_0^2}{2a_0},\ \ \ \ b_1=\frac{b_0}{2a_0},\\
 &&R_1=R_0+\frac{b_0(S_0+R_0)}{2a_0}\ \ S_0=S_0+\frac{b_0(S_0+R_0)}{2a_0}\ \ \Omega_1=\frac{(R_0+S_0)^2}{8a_0}
\label{eq:cof_grp_1}
\end{eqnarray}
Integration is first done with respect to $x_1$, we have
\[
 A_2=A_1\left(\frac{\ui\pi\hbar}{2a_0}\right)^{1/2}\exp\left(-\frac{\ui}{\hbar}\Omega_1\right)
\]
\[
  \phi_2=\sum_{j=2}^{N-1}\left[a_0(x_{j+1}^2+x_j^2)-2b_0x_{j+1}x_j\right]+a_1(x_2^2+x_0^2)-2b_1x_2x_0-R_1x_2-S_1x_0
\]
\begin{eqnarray}
  \phi_k&=&\sum_{j=k}^{N-1}\left[a_0(x_{j+1}^2+x_j^2)-2b_0x_{j+1}x_j\right]+a_{k-1}(x_k^2+x_0^2)-2b_{k-1}x_kx_0-R_{k-1}x_k-S_{k-1}x_0 \nonumber\\
&=&\sum_{j=k+1}^{N-1}\left[a_0(x_{j+1}^2+x_j^2)-2b_0x_{j+1}x_j\right]+\alpha_k,
\end{eqnarray}
where 
\begin{eqnarray*}
 \alpha_k&=&a_0(x_{k+1}^2+x_k^2)-2b_0x_{k+1}x_k-R_0X_{k+1}-S_0x_k\\
&&+a_{k-1}(x_k^2+x_0^2)-2b_{k-1}x_kx_0-R_{k-1}X_k-S_{k-1}x_0\\
&=&a_0x_{k+1}^2+a_{k-1}x_0^2-R_0X_{k+1}-S_{k-1}x_0+
\frac{1}{a_0+a_{k-1}}\left[x_k-\frac{(b_0x_{k+1}+b_{k-1}x_0)+(R_{k-1}+S_0)/2}{a_0+a_{k-1}}\right]^2\\
&&-\frac{\left[(b_0x_{k+1}+b_{k-1}x_0)+(R_{k-1}+S_0)/2\right]^2}{a_0+a_{k-1}}
\end{eqnarray*}
Integration is first done with respect to $x_k$, we have
\[
 A_{k+1}=A_k\sqrt{\frac{\ui\pi\hbar}{a_0+a_{k-1}}}\exp\left(-\frac{\ui}{\hbar}\Omega_k\right)
\]
\begin{eqnarray}
 \phi_{k+1}&=&\sum_{j=k+1}^{N-1}\left[a_0(x_{j+1}^2+x_j^2)-2b_0x_{j+1}x_j\right]+a_0x_{k+1}^2+a_{k-1}x_0^2-R_0X_{k+1}-S_{k-1}x_0\nonumber\\
&&-\frac{\left[(b_0x_{k+1}+b_{k-1}x_0)+(R_{k-1}+S_0)/2\right]^2}{a_0+a_{k-1}}\nonumber\\
&=&\sum_{j=k+1}^{N-1}\left[a_0(x_{j+1}^2+x_j^2)-2b_0x_{j+1}x_j\right]\nonumber\\
&&+a_kx_{k+1}^2+a_k'x_0^2-2b_kx_{k+1}x_0-R_kx_{k+1}-S_kx_0-\Omega_k,
\label{eq:phi_k+1}
\end{eqnarray}
where
\begin{eqnarray}
 &&a_k=a_0-\frac{b_0^2}{a_0+a_{k-1}},\ \ \ \ a_k'=a_{k-1}-\frac{b_{k-1}^2}{a_0+a_{k-1}}\ \ \ \ b_k=\frac{b_0b_{k-1}}{a_0+a_{k-1}},\label{eq:cof_grp_k_1}\\
 &&R_k=R_0+\frac{(R_{k-1}+S_0)b_0}{a_0+a_{k-1}},\ \ \ \ S_k=S_{k-1}+\frac{(R_{k-1}+S_0)b_{k-1}}{a_0+a_{k-1}},\ \ \ \ \Omega_k=\frac{(R_{k-1}+S_0)^2}{4(a_0+a_{k-1})}
\label{eq:cof_grp_k_2}
\end{eqnarray}
If $a_k=a_k'$, i.e.
\begin{equation}
 a_{k-1}^2=b_{k-1}^2+a_0^2-b_0^2,
\label{eq:consist_1}
\end{equation}
then the form of $\phi_{k+1}$ is same as the form of $\phi_{k}$. Eq. (\ref{eq:consist_1}) is tenable as $k=1,k=2$, 
because from Eq. (\ref{eq:cof_grp_1}) can be derived
\[
 a_1^2-b_1^2=\left(a_0-\frac{b_0^2}{2a_0}\right)^2-\left(\frac{b_0^2}{2a_0}\right)^2=a_0^2-b_0^2.
\]
Eq. (\ref{eq:phi_k+1}) and Eq. (\ref{eq:cof_grp_k_1}) can be proved by mathematical induction for $k=1,2,3, \dots$.
\begin{eqnarray*}
A_N&=&A_{N-1}\sqrt{\frac{\ui\pi\hbar}{a_0+a_{N-2}}}\exp\left(-\frac{\ui}{\hbar}\Omega_{N-1}\right)
=\left(\frac{1}{2\pi \ui\hbar\epsilon}\right)^{N/2}\exp{\left(\sum_{k=1}^{N-1}-\frac{\ui}{h}\Omega_{k}\right)}\prod_{k=1}^{N-1}
\sqrt{\frac{\ui\pi\hbar}{a_0+a_{k-1}}}\nonumber\\
\phi_N&=&a_{N-1}(x_N^2+x_0^2)-2b_{N-1}x_Nx_0-R_{N-1}x_N-S_{N-1}x_0
\end{eqnarray*}
Substituting the equation above into Eq. (\ref{eq:dp_prpgt_1}), we have
\begin{eqnarray}
 K(x_b,T;x_a,0)&=&\lim_{\substack{N \rightarrow \infty\\ \epsilon\rightarrow 0}}
\left(\frac{1}{2\pi \ui\hbar\epsilon}\right)^{1/2}\left[\prod_{k=1}^{N-1}\left(\frac{1}{2\epsilon}\frac{1}{a_0+a_{k-1}}\right)\right]^{1/2}
\exp{\left(\sum_{k=1}^{N-1}-\frac{\ui}{h}\Omega_{k-1}\right)}\nonumber\\
&&\times \exp{\left\lbrace \frac{\ui}{h}\left[a_{N-1}(x_N^2+x_0^2)-2b_{N-1}x_Nx_0-R_{N-1}x_N-S_{N-1}x_0\right]\right\rbrace}
\label{eq:dp_prpgt_2}
\end{eqnarray}
The coefficients $a_k,b_k$ of Eq. (\ref{eq:dp_prpgt_2}) can be derived from the following recursion formulas:
\begin{equation}
a_0=b_0\left(1+2\left(\frac{\kappa\epsilon}{2}\right)^2\right),\ \ \ \ b_0=\frac{1}{2\epsilon} 
\label{eq:ab_0}
\end{equation}
\begin{equation}
a_{k-1}=(b_{k-1}^2+a_0^2-b_0^2)^{1/2},\ \ \ \ b_k=\frac{b_0b_{k-1}}{a_0+a_{k-1}}
\label{eq:ab_k}
\end{equation}

In order to obtain the result of Eq. (\ref{eq:dp_prpgt_2}) with $N\rightarrow \infty$, one must obtain the limit of $a_k$, $b_k$, $R_k$, $S_k$, $\Omega_k$. 
Let us consider $a_k,\ \ b_k$ first. As $N\rightarrow \infty$, we have
\begin{equation}
 \frac{1}{2}\kappa \epsilon=\sinh{\frac{1}{2}\kappa \epsilon}.
\label{eq:eqvinfs}
\end{equation}
Substituting Eq. (\ref{eq:eqvinfs}) into Eq. (\ref{eq:ab_0}), we have
\[
 a_0=b_0\left(1+2\sinh^2{\frac{1}{2}\kappa \epsilon}\right)=b_0\cosh{\kappa \epsilon}
\]
Substituting the equation above into Eq. (\ref{eq:ab_k}), we have
\begin{equation}
 \frac{1}{b_k}=\frac{\cosh{\kappa\epsilon}}{b_{k-1}}+\frac{1}{b_0}\sqrt{1+\frac{b_0^2}{b_{k-1}^2}\sinh \kappa \epsilon}
\label{eq:recp_bk}
\end{equation}
Let $k=1$, Eq. (\ref{eq:recp_bk}) becomes
\begin{equation}
 \frac{1}{b_1}=\frac{1}{b_0}(\cosh \kappa\epsilon+\sqrt{1-\sinh^2\kappa\epsilon})=\frac{2\cosh{\kappa \epsilon}}{b_0}
=\frac{\sinh{2\kappa \epsilon}}{b_0\sinh \kappa \epsilon}.
\end{equation}
Let $k=2$, we have
\begin{equation*}
 \frac{1}{b_2}=\frac{\cosh \kappa\epsilon}{b_1}+\frac{1}{b_0}\sqrt{1+\frac{b_0^2}{b_1^2}\sinh^2 \kappa \epsilon}
=\frac{\sinh 3\kappa\epsilon}{b_1 \sinh 2\kappa \epsilon}=\frac{\sinh 3\kappa\epsilon}{b_0 \sinh \kappa \epsilon}
\end{equation*}
The rest may be deduced by analogy, we have
\[
 \frac{1}{b_{k-1}}=\frac{\sinh k\kappa\epsilon }{b_0\sinh \kappa \epsilon}
\]
The equation above can be proved by recursion method. Substituting the equation above into Eq. (\ref{eq:recp_bk}), we have
\begin{eqnarray*}
 \frac{1}{b_{k-1}}&=&\frac{\cosh \kappa\epsilon\sinh k\kappa\epsilon}{b_0\sinh \kappa\epsilon}
+\frac{1}{b_0}\sqrt{1+b_0^2\left(\frac{\sinh k\kappa\epsilon}{b_0 \sinh \kappa \epsilon}\right)^2 \sinh^2\kappa\epsilon}\\
&=&\frac{\sinh(k+1)\kappa\epsilon}{b_0\sinh \kappa\epsilon}
\end{eqnarray*}
Hence
\begin{equation}
 b_k=\frac{1}{2\epsilon}\frac{\sinh \kappa\epsilon}{\sinh (k+1)\kappa\epsilon}
\label{eq:bk_rp}
\end{equation}
\begin{equation}
 a_k=(b_k^2+a_0^2-b_0^2)^{1/2}=b_k\sqrt{1+\frac{b_0^2}{b_k^2}\sinh^2\kappa\epsilon}
=\frac{\sinh \kappa\epsilon}{2\epsilon}\frac{\cosh (k+1)\kappa\epsilon}{\sinh (k+1)\kappa \epsilon}
\label{eq:ak_rp}
\end{equation}
\begin{eqnarray}
 a_0+a_{k-1}&=&\frac{1}{2\epsilon}\left(cosh\kappa\epsilon+\sinh \kappa\epsilon\frac{\cosh k\kappa\epsilon}{\sinh k\kappa \epsilon}\right)\nonumber\\
&=&\frac{1}{2\epsilon}\frac{\sinh(k+1)\kappa\epsilon}{\sin \kappa\epsilon}
\label{eq:a0ak-1}
\end{eqnarray}
Substituting the equation above into Eq. (\ref{eq:cof_grp_k_2}) and let $k=1$, we have
\begin{eqnarray*}
R_1&=&R_0+\frac{(R_0+S_0)b_0}{a_0+a_0}=\frac{\kappa^2\Lambda\epsilon}{2}\left(1+\frac{1}{\cosh\kappa\epsilon}\right)\\
S_1&=&S_0+\frac{(R_0+S_0)b_0}{a_0+a_0}=\frac{\kappa^2\Lambda\epsilon}{2}\left(1+\frac{1}{\cosh\kappa\epsilon}\right)
\end{eqnarray*}
Let $k=2$ and substitute Eq. (\ref{eq:a0ak-1}) into Eq. (\ref{eq:cof_grp_k_2}), we have
\begin{eqnarray*}
 R_2&=R&_0+\frac{(R_1+S_0)b_0}{a_0+a_1}
=\frac{\kappa^2\Lambda\epsilon}{2}\left[1+
\frac{1}{\cosh \kappa\epsilon}\frac{(2\cosh\kappa\epsilon+1)\sinh2\kappa\epsilon}{\sinh 3\kappa\epsilon}\right]\\
S_2&=&S_1+\frac{R_1+S_0}{a_0+a_1}b_1=\frac{\kappa^2\Lambda\epsilon}{2}\left\lbrace 1+
\frac{1}{\cosh\kappa\epsilon}\left[1+\frac{(2\cosh\kappa\epsilon+1)\sinh\kappa\epsilon}{\sinh3\kappa\epsilon}\right]\right\rbrace 
\end{eqnarray*}
For $k=3$, we have
\begin{eqnarray*}
 R_3&=&R_0+\frac{R_2+S_0}{a_0+a_2}b_0=\frac{\kappa^2\Lambda\epsilon}{2}\left[1+\frac{1}{\cosh \kappa\epsilon}
\frac{2\cosh\kappa\epsilon\sinh 3\kappa\epsilon+(2\cosh\kappa\epsilon+1)\sinh 2\kappa\epsilon}{\sinh 4\kappa\epsilon}\right]\\
S_3&=&S_2+\frac{R_2+S_0}{a_0+a_2}b_2\\
&=&\frac{\kappa^2\Lambda\epsilon}{2}\left\lbrace 1+\frac{1}{\cosh\kappa\epsilon}
\left[1+(2\cosh(\kappa\epsilon+1)\frac{\sinh\kappa\epsilon}{\sinh 3\kappa\epsilon} \right.\right. \\ 
&&\left.\left.+\frac{2\cosh\kappa\epsilon\sinh3\kappa\epsilon +(2\cosh\kappa\epsilon+1)\sinh2\kappa\epsilon}{\sinh3\kappa\epsilon}\frac{\sinh\kappa\epsilon}{\sinh4\kappa\epsilon}
\right]\right\rbrace
\end{eqnarray*}
We can suppose
\begin{equation}
 R_k=\frac{\kappa^2 \Lambda\epsilon}{2}\left[1
+\frac{1}{\cosh \kappa\epsilon}\frac{\sum_{j=1}^k 2\cosh\kappa\epsilon\sinh j\kappa\epsilon}{\sinh (k+1)\kappa\epsilon}\right]
\label{eq:eq_Rk}
\end{equation}
\begin{equation}
 S_k=\frac{\kappa^2\Lambda\epsilon}{2}\left\lbrace 1+\frac{1}{\cosh\kappa\epsilon}
\sum_{j=1}^k\frac{\sinh \kappa\epsilon}{\sinh (j+1)\kappa\epsilon}
\left[\frac{\sum_{l=1}^j 2\cosh\kappa\epsilon\sinh l\kappa\epsilon}{\sinh j\kappa\epsilon}\right]\right\rbrace
\label{eq:eq_Sk}
\end{equation}
Substituting Eq. (\ref{eq:eq_Rk}) into the last term in Eq. (\ref{eq:cof_grp_k_2}), we have
\begin{equation}
 \Omega_k=\frac{\kappa^3\Lambda^2\epsilon^3}{4}\frac{\sinh \kappa\epsilon}{\sinh (k+1)\kappa\epsilon}\left[2+
\frac{1}{\cosh \kappa\epsilon}\frac{\sum_{j=1}^{k-1} 2\cosh\kappa\epsilon\sinh j\kappa\epsilon}{\sinh k\kappa\epsilon}\right]^2
\end{equation}
Therefore, we have the limit of $R_k,S_k,\Omega_k$
\begin{equation}
 \lim_{N\rightarrow\infty}R_N=\frac{\kappa^2\Lambda}{2}\frac{\ue^{2\kappa T}-2\ue^{\kappa T}+1}{\kappa \ue^{\kappa T}\sinh \kappa T}
=\kappa\Lambda\frac{\ue^{\kappa T}-1}{\ue^{\kappa T}+1}
\label{eq:lim_Rk}
\end{equation}
\begin{equation}
 \lim_{N\rightarrow\infty}S_N=\kappa\Lambda\frac{\ue^{\kappa T}-1}{\ue^{\kappa T}+1}
\label{eq:lim_Sk}
\end{equation}
%\[
% \lim_{N\rightarrow\infty}\sum_{k=1}^{N-1}\Omega_{k-1}=1/2\, \left( -2\,\ln  \left( 2 \right) +2\,\ln  \left( 1+{{\rm e}^{
%\kappa\,T}} \right) -\kappa\,T \right) \Lambda
%\]
\begin{equation}
 \lim_{N\rightarrow\infty}\sum_{k=1}^{N-1}\Omega_{k-1}=0
\label{eq:lim_Omg}
\end{equation}
%\begin{eqnarray}
%  K(x_b,t_b;x_a,t_a)&=&\lim_{\substack{N \rightarrow \infty\\ \epsilon\rightarrow 0}}
%\left(\frac{1}{2\pi i\hbar\epsilon}\right)^{1/2}\left[\prod_{k=1}^{N-1}\left(\frac{1}{2\epsilon}\frac{1}{a_0+a_{k-1}}\right)\right]^{1/2}
%exp{\left(\sum_{k=1}^{N-1}-\frac{i}{h}\Omega_{k-1}\right)}\nonumber\\
%&&\times \exp{\left\lbrace \frac{i}{h}\left[a_{N-1}(x_N^2+x_0^2)-2b_{N-1}x_Nx_0-R_{N-1}x_N-S_{N-1}x_0\right]\right\rbrace}\nonumber\\
%&=&\lim_{\substack{N \rightarrow \infty\\ \epsilon\rightarrow 0}}\left(\frac{1}{2\pi i\hbar\epsilon}\right)^{1/2}
%\left[\prod_{k=1}^{N-1}\left(\frac{\sinh \kappa\epsilon}{\sinh (k+1)\kappa\epsilon}\right)\right]^{1/2}
%\exp\left(\lim_{N\rightarrow\infty}\sum_{k=1}^{N-1}\Omega_{k-1}\right)\times\nonumber\\
%&&\exp\left\lbrace\frac{i}{\hbar}\left[\frac{\sinh \kappa\epsilon}{2\epsilon}\frac{\cosh N\kappa\epsilon}{\sinh N\kappa \epsilon}
%(x_N^2+x_0^2)-\frac{1}{\epsilon}\frac{\sinh \kappa\epsilon}{\sinh (k+1)\kappa\epsilon}x_Nx_0 \right.\right.\nonumber\\ 
%&&\left.\left.
%-\lim_{N\rightarrow\infty}R_N x_N-\lim_{N\rightarrow\infty}S_Nx_0\right]\right\rbrace\nonumber\\
%&=&\left(\frac{\kappa}{2\pi i\hbar\sinh \kappa T}\right)^{1/2}left(\cosh\frac{\kappa T}{2}\right)^\Lambda
%\exp \left\lbrace\frac{i}{\hbar}\frac{\kappa}{2\sinh \kappa T} \right.\nonumber\\
%&&\left.\left[\cosh \kappa T(x_b^2+x_a^2)-2x_bx_a\right]
%-\kappa\Lambda\frac{e^{\kappa T}-1}{e^{\kappa T}+1}(x_b+x_a) 
%\right\rbrace
%\end{eqnarray}
Substituting Eq. (\ref{eq:ak_rp}) and Eq. (\ref{eq:bk_rp}), Eq. (\ref{eq:a0ak-1}), Eq. (\ref{eq:lim_Rk}), Eq. (\ref{eq:lim_Rk}), Eq. (\ref{eq:lim_Omg}) 
into Eq. (\ref{eq:dp_prpgt_1}), we have
\begin{eqnarray}
  K(x_b,T;x_a,0)&=&\lim_{\substack{N \rightarrow \infty\\ \epsilon\rightarrow 0}}
\left(\frac{1}{2\pi \ui\hbar\epsilon}\right)^{1/2}\left[\prod_{k=1}^{N-1}\left(\frac{1}{2\epsilon}\frac{1}{a_0+a_{k-1}}\right)\right]^{1/2}
\exp{\left(\sum_{k=1}^{N-1}-\frac{\ui}{h}\Omega_{k-1}\right)}\nonumber\\
&&\times \exp{\left\lbrace \frac{\ui}{h}\left[a_{N-1}(x_N^2+x_0^2)-2b_{N-1}x_Nx_0-R_{N-1}x_N-S_{N-1}x_0\right]\right\rbrace}\nonumber\\
&=&\lim_{\substack{N \rightarrow \infty\\ \epsilon\rightarrow 0}}\left(\frac{1}{2\pi \ui\hbar\epsilon}\right)^{1/2}
\left[\prod_{k=1}^{N-1}\left(\frac{\sinh \kappa\epsilon}{\sinh (k+1)\kappa\epsilon}\right)\right]^{1/2}
\exp\left(\lim_{N\rightarrow\infty}\sum_{k=1}^{N-1}\Omega_{k-1}\right)\times\nonumber\\
&&\exp\left\lbrace\frac{\ui}{\hbar}\left[\frac{\sinh \kappa\epsilon}{2\epsilon}\frac{\cosh N\kappa\epsilon}{\sinh N\kappa \epsilon}
(x_N^2+x_0^2)-\frac{1}{\epsilon}\frac{\sinh \kappa\epsilon}{\sinh (k+1)\kappa\epsilon}x_Nx_0 \right.\right.\nonumber\\ 
&&\left.\left.
-\lim_{N\rightarrow\infty}R_N x_N-\lim_{N\rightarrow\infty}S_Nx_0\right]\right\rbrace\nonumber\\
&=&\left(\frac{\kappa}{2\pi \ui\hbar\sinh \kappa T}\right)^{1/2}
\exp \left\lbrace\frac{\ui}{\hbar}\left\lbrace\frac{\kappa}{2\sinh \kappa T} 
\left[\cosh \kappa T(x_b^2+x_a^2)-2x_bx_a\right]\right.\right.\\
&&\left.\left.-\kappa\Lambda\frac{\ue^{\kappa T}-1}{\ue^{\kappa T}+1}(x_b+x_a) 
\right\rbrace\right\rbrace
\label{eq:eq:dp_prpgt_2}
\end{eqnarray}
Substituting Eq. (\ref{eq:bndcondt}) into the Equation above, we have
\begin{eqnarray}
  K(x_b,T;x_a,0)&=&\left(\frac{\kappa}{2\pi \ui\hbar\sinh \kappa T}\right)^{1/2}
\exp \left\lbrace\frac{\ui}{\hbar}\left\lbrace\frac{\kappa}{2\sinh \kappa T} 
\left[\cosh \kappa T(x_b^2+x_a^2)-2x_bx_a\right]\right.\right.\nonumber\\
&&\left.\left.-\kappa\left(x_a-\frac{x_b-x_a}{\ue^{-\kappa T}-1}\right)\frac{\ue^{\kappa T}-1}{\ue^{\kappa T}+1}(x_b+x_a) 
\right\rbrace\right\rbrace \nonumber\\
&=&\left(\frac{\kappa}{2\pi \ui\hbar\sinh \kappa T}\right)^{1/2}
\exp \left\lbrace\frac{\ui}{\hbar}\left[\frac{\kappa}{2\tanh \kappa T} 
(x_b^2+x_a^2)-\frac{\kappa x_bx_a}{\sinh \kappa T}\right.\right.\nonumber\\
&&\left.\left.-\kappa\left(x_a-\frac{x_b-x_a}{\ue^{-\kappa T}-1}\right)\tanh \frac{\kappa T}{2}(x_b+x_a) 
\right]\right\rbrace.
\label{eq:dp_prpgt_3}
\end{eqnarray}
It is evident that the propagator above tends to the free Schr\"{o}dinger propagator in the limit $\kappa\rightarrow 0$.

Finally we perform an analysis of the time evolution in terms of the quantum propagator (\ref{eq:dp_prpgt_3}).Suppose
that at the initial time we have a Gaussian wave-packet 
\[
 \psi_{0}(q)\propto \exp\left(-\theta_0 q^2+\frac{\ui}{\hbar}v_0 q\right),
\]
, which describes a unit mass particle localized in a
neighborhood of the point $q= 0$ with an initial velocity  $v_0$. 
At a later time t, the system under consideration
will be characterized by the propagated wave-packet distribution, i.e.
\begin{equation}
 \psi_{T}(x)=\int_{-\infty}^{\infty}K(x,T;q,0)\psi_{0}(q)\dif q 
\propto \exp \left\lbrace -\theta_1 (x-\langle x\rangle)^2+\frac{\ui}{\hbar}\theta_2 q^2 \right\rbrace.
\label{eq:gaussian_wave}
\end{equation}
The direct calculation shows that the Wave-function $\psi_{T}(x)$ still remains Gaussian. The mean value of position is
\begin{equation}
 \langle x\rangle=v_0\frac{\tanh\kappa T}{\kappa}
\label{eq:mx_lg}
\end{equation}
The mean value of velocity of the wave packet
\begin{equation}
 \langle v\rangle=\frac{2k\,\tanh\left( \frac{k\,T}{2}\right) }{{\ue}^{-k\,T}-1}\langle x\rangle+v_0
\label{eq:mv_lg}
\end{equation}
The coefficient $\theta_1$ in Eq. (\ref{eq:gaussian_wave})
\begin{equation}
 \theta_1=\frac{\frac{\tanh^2(\kappa*T)}{\hbar^2 \kappa^2}}
{4\left[ \alpha_0-\frac{\ui}{\hbar}\left(\frac{\kappa}{2\tanh \kappa T}-\kappa \tanh \frac{\kappa T}{2}-\kappa\frac{\tanh \frac{\kappa T}{2}}{\ue^{-\kappa T}-1}\right)\right]}
\label{eq:alpha_lg}
\end{equation}

Let us compare our result with the results obtained by Kochan's approach\cite{PhysRevA.81.022112,2009IJMPA..24.5319K} and Caldirola-Kanai's method\cite{Das2005}, 
Das's method\cite{Das2005}.
 \begin{figure}
 \includegraphics{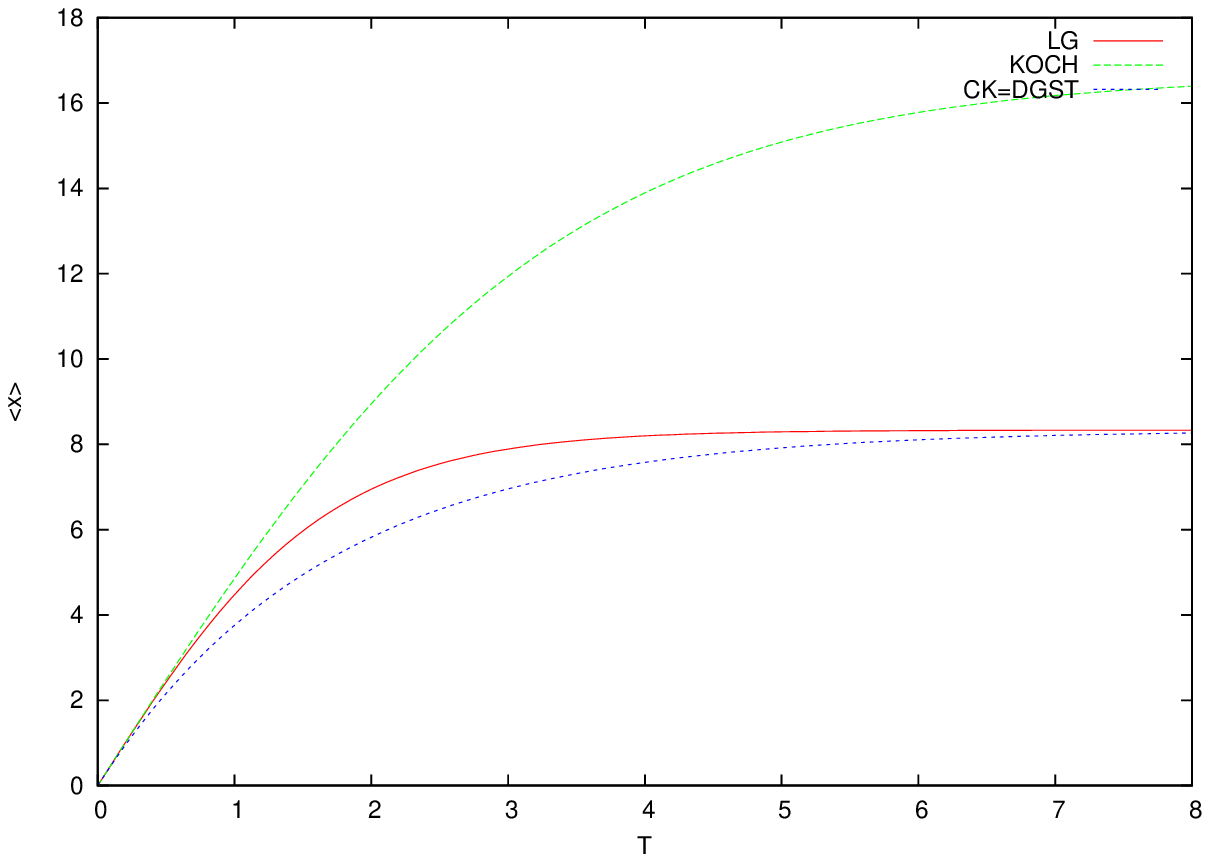}
\includegraphics{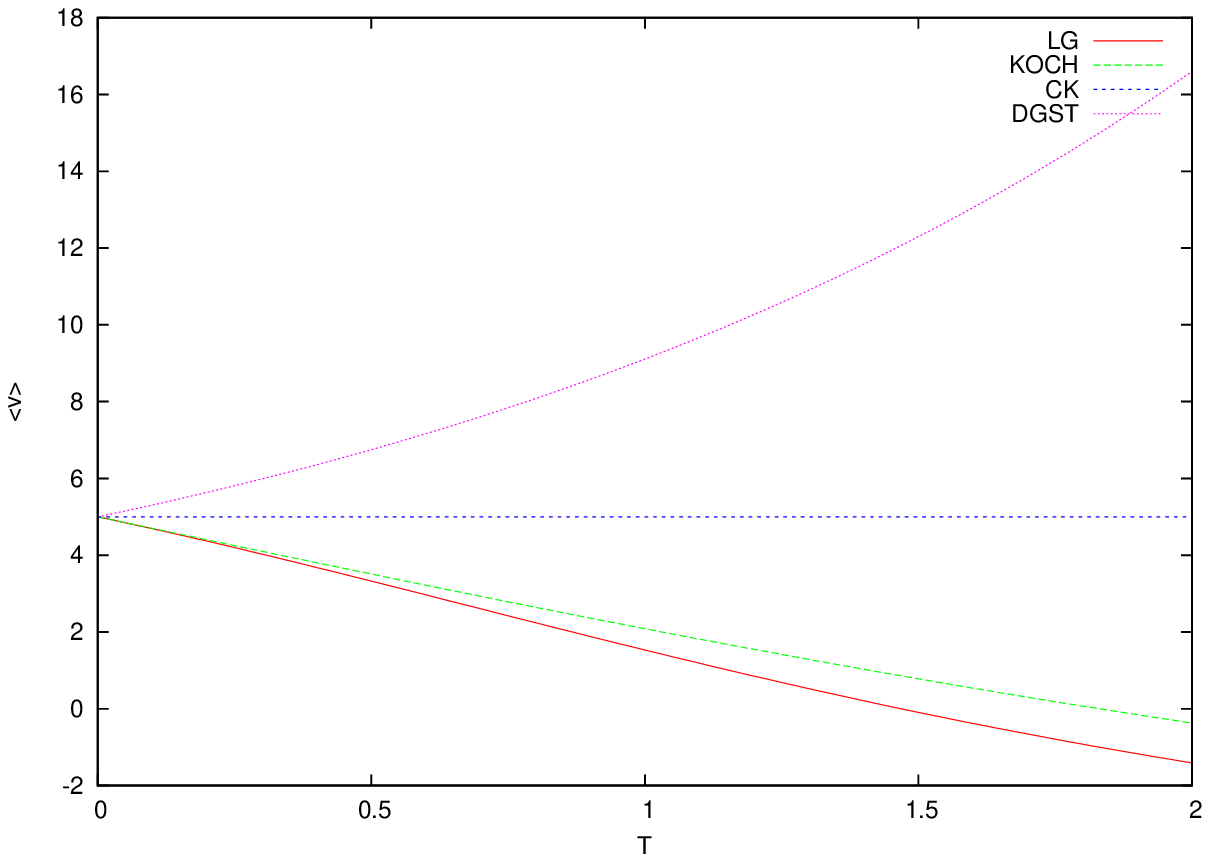}
 \caption{Comparison of the mean values of position and velocity obtained by various methods\label{fig:comp}}
 \end{figure}
The results obtained by Kochan's (abbr. Koch) approach\cite{Kochan2010219} are
\begin{eqnarray*}
\langle x\rangle_{Koch}&=&2v_0\frac{\tanh\frac{\kappa T}{2}}{\kappa} \\
\langle v\rangle_{Koch}&=&\frac{2v_0}{1+\ue^{-\kappa T}}-\frac{3}{2}\kappa \langle x\rangle\\
\theta_{1_{Koch}}&=&\frac{\frac{\kappa^2}{16\hbar^2\tanh^2\left\lbrace0.5\kappa T\right\rangle}}
{\alpha_0-\ui \frac{\kappa(3-\ue^{-\kappa T})}{4\hbar(1-\ue^{\kappa T})}}.
\end{eqnarray*}
The results obtained by Caldirola-Kanai (abbr. CK) method are
\begin{eqnarray*}
 \langle x\rangle_{CK}&=&\frac{v_0}{\kappa}(1-\ue^{-\kappa T})\\
\langle v\rangle_{CK}&=&v_0\\
\theta_{1_{CK}}&=&\frac{\frac{\kappa^2}{4\hbar^2(1-\ue^{-\kappa T})^2}}{\alpha_0-\ui\frac{\kappa }{2\hbar(1-\ue^{-\kappa T})}}.
\end{eqnarray*}
The results obtained by Das's (abbr. DGST) method are
\begin{eqnarray*}
 \langle x\rangle_{DGST}=\frac{v_0}{\kappa}(1-\ue^{-\kappa T})\\
 \langle v\rangle_{DGST}=v_0\ue^{\kappa T}
 \theta_{1_{DGST}}&=&\frac{\frac{\kappa^2}{4\hbar^2(1-\ue^{-\kappa T})^2}}{\alpha_0-\ui\frac{\kappa }{2\hbar(1-\ue^{-\kappa T})}}.
\end{eqnarray*}
The comparison among these results above is illustrated in Fig. \ref{fig:comp}, 
where $q=0 \mathrm{m}, \ \ v_0=5 \mathrm{ms^{-1}}, \ \ \kappa=0.6 \mathrm{s}^{-1}$
. The curve obtained by our method is labeled with 'LG'.
It is clear that the expectation value of velocity obtained by our method and Kochan's method look very similar. The expectation value 
of position obtained by our method is quite different from that obtained by Kochan's method, but with time evolution, $\langle x \rangle$ obtained by our 
method approaches $\langle x\rangle_{CK}$. The reason is 
\[
 \lim_{T\rightarrow \infty}{\langle x \rangle}=\lim_{T\rightarrow \infty}{\langle x \rangle_{CK}}=\frac{v_0}{\kappa},\ \ \ \
 \lim_{T\rightarrow \infty}{\langle x \rangle}_{Koch}=\frac{2v_0}{\kappa}.
\]
Because our averaged velocity $\langle v\rangle$ and Kochan's averaged velocity posses a similar advantage that they decrease with respect to 
time translations, also possess similar restriction that our averaged velocity is reliable in the time interval $\left[0,\ln(\sqrt 2+1)/\kappa \right]$ and
Kochan's reliable interval is $\left[0,\ln3/\kappa \right]$. Exactly at the moment $\ln(\sqrt 2+1)/\kappa$, our averaged velocity becomes zero and going over, 
it would produce a negative (nonphysical) mean velocity.

\section{Conclusion}
Kochan\cite{2009IJMPA..24.5319K} have compared his results with other results obtained by the method of Caldirola-Kanai and other methods. 
Because the aforementioned similarity, our approach possesses the advantage of Kochan's approach, but the position obtained by our approach likes that obtained 
by CK-method. Since both Kochan's approach and our approach are based ultimately on the classical history, this similarity exists between the approaches.
For quantization of more complicated damping mechanical systems, one can utilize numerical method to obtain classical trajectory, then
 construct approximation of the Lagrangian density.


\begin{thebibliography}{10}%
\makeatletter
\providecommand \@ifxundefined [1]{%
 \ifx #1\undefined \expandafter \@firstoftwo
 \else \expandafter \@secondoftwo
\fi
}%
\providecommand \@ifnum [1]{%
 \ifnum #1\expandafter \@firstoftwo
 \else \expandafter \@secondoftwo
\fi
}%
\providecommand \enquote [1]{``#1''}%
\providecommand \bibnamefont  [1]{#1}%
\providecommand \bibfnamefont [1]{#1}%
\providecommand \citenamefont [1]{#1}%
\providecommand\href[0]{\@sanitize\@href}%
\providecommand\@href[1]{\endgroup\@@startlink{#1}\endgroup\@@href}%
\providecommand\@@href[1]{#1\@@endlink}%
\providecommand \@sanitize [0]{\begingroup\catcode`\&12\catcode`\#12\relax}%
\@ifxundefined \pdfoutput {\@firstoftwo}{%
 \@ifnum{\z@=\pdfoutput}{\@firstoftwo}{\@secondoftwo}%
}{%
 \providecommand\@@startlink[1]{\leavevmode}%
 \providecommand\@@endlink[0]{}%
}{%
 \providecommand\@@startlink[1]{%
  \leavevmode
  \pdfstartlink
   attr{/Border[0 0 1 ]/H/I/C[0 1 1]}%
   user{/Subtype/Link/A<</Type/Action/S/URI/URI(#1)>>}%
  \relax
 }%
 \providecommand\@@endlink[0]{\pdfendlink}%
}%
\providecommand \url  [0]{\begingroup\@sanitize \@url }%
\providecommand \@url [1]{\endgroup\@href {#1}{\urlprefix}}%
\providecommand \urlprefix [0]{URL }%
\providecommand \Eprint[0]{\href }%
\@ifxundefined \urlstyle {%
  \providecommand \doi [1]{doi:\discretionary{}{}{}#1}%
}{%
  \providecommand \doi [0]{doi:\discretionary{}{}{}\begingroup
  \urlstyle{rm}\Url }%
}%
\providecommand \doibase [0]{http://dx.doi.org/}%
\providecommand \Doi[1]{\href{\doibase#1}}%
\providecommand \selectlanguage [0]{\@gobble}%
\providecommand \bibinfo [0]{\@secondoftwo}%
\providecommand \bibfield [0]{\@secondoftwo}%
\providecommand \translation [1]{[#1]}%
\providecommand \BibitemOpen[0]{}%
\providecommand \bibitemStop [0]{}%
\providecommand \bibitemNoStop [0]{.\EOS\space}%
\providecommand \EOS [0]{\spacefactor3000\relax}%
\providecommand \BibitemShut [1]{\csname bibitem#1\endcsname}%
%</preamble>
\bibitem{2005RvMaP..17..391A}%
  \BibitemOpen
  \bibfield{author}{%
  \bibinfo {author} {\bibfnamefont{S.~T.}\ \bibnamefont{{Ali}}}\ and\ \bibinfo
  {author} {\bibfnamefont{M.}~\bibnamefont{{Engli{\v s}}}},\ }%
  \bibfield{title}{%
  \enquote{\bibinfo {title} {{Quantization Methods:}},}\ }%
  \bibfield{journal}{%
  \Doi{10.1142/S0129055X05002376}{\bibinfo {journal} {Reviews in Mathematical
  Physics}}\ }%
  \textbf{\bibinfo {volume} {17}},\ \bibinfo {pages} {391--490} (\bibinfo
  {year} {2005})\BibitemShut{NoStop}%
\bibitem{1986AmJPh..54..273H}%
  \BibitemOpen
  \bibfield{author}{%
  \bibinfo {author} {\bibfnamefont{L.}~\bibnamefont{{Herrera}}}, \bibinfo
  {author} {\bibfnamefont{L.}~\bibnamefont{{N{\'u}{\~n}ez}}}, \bibinfo {author}
  {\bibfnamefont{A.}~\bibnamefont{{Pati{\~n}o}}},\ and\ \bibinfo {author}
  {\bibfnamefont{H.}~\bibnamefont{{Rago}}},\ }%
  \bibfield{title}{%
  \enquote{\bibinfo {title} {{A variational principle and the classical and
  quantum mechanics of the damped harmonic oscillator}},}\ }%
  \bibfield{journal}{%
  \Doi{10.1119/1.14644}{\bibinfo {journal} {American Journal of Physics}}\ }%
  \textbf{\bibinfo {volume} {54}},\ \bibinfo {pages} {273--277} (\bibinfo
  {month} {Mar.}\ \bibinfo {year} {1986})\BibitemShut{NoStop}%
\bibitem{springerlink:10.1007/s10773-006-9064-9}%
  \BibitemOpen
  \bibfield{author}{%
  \bibinfo {author} {\bibfnamefont{G.}~\bibnamefont{López}}\ and\ \bibinfo
  {author} {\bibfnamefont{P.}~\bibnamefont{López}},\ }%
  \bibfield{title}{%
  \enquote{\bibinfo {title} {Velocity quantization approach of the
  one-dimensional dissipative harmonic oscillator},}\ }%
  \bibfield{journal}{%
  \bibinfo {journal} {International Journal of Theoretical Physics}\ }%
  \textbf{\bibinfo {volume} {45}},\ \bibinfo {pages} {734--742} (\bibinfo
  {year} {2006}),\ ISSN \bibinfo {issn} {0020-7748},\ \bibinfo {note}
  {10.1007/s10773-006-9064-9},\
  \url{http://dx.doi.org/10.1007/s10773-006-9064-9}\BibitemShut{NoStop}%
\bibitem{Chruscinski2006854}%
  \BibitemOpen
  \bibfield{author}{%
  \bibinfo {author} {\bibfnamefont{Dariusz}\ \bibnamefont{Chruscinski}}\ and\
  \bibinfo {author} {\bibfnamefont{Jacek}\ \bibnamefont{Jurkowski}},\ }%
  \bibfield{title}{%
  \enquote{\bibinfo {title} {Quantum damped oscillator i: Dissipation and
  resonances},}\ }%
  \bibfield{journal}{%
  \Doi{DOI: 10.1016/j.aop.2005.11.004}{\bibinfo {journal} {Annals of Physics}}\
  }%
  \textbf{\bibinfo {volume} {321}},\ \bibinfo {pages} {854 -- 874} (\bibinfo
  {year} {2006}),\ ISSN \bibinfo {issn} {0003-4916},\
  \url{http://www.sciencedirect.com/science/article/B6WB1-4HWXP35-3/2/16e9c1be%
2f24e1d227c355fad3a95610}\BibitemShut{NoStop}%
\bibitem{Blasone2004354}%
  \BibitemOpen
  \bibfield{author}{%
  \bibinfo {author} {\bibfnamefont{Massimo}\ \bibnamefont{Blasone}}\ and\
  \bibinfo {author} {\bibfnamefont{Petr}\ \bibnamefont{Jizba}},\ }%
  \bibfield{title}{%
  \enquote{\bibinfo {title} {Bateman's dual system revisited: quantization,
  geometric phase and relation with the ground-state energy of the linear
  harmonic oscillator},}\ }%
  \bibfield{journal}{%
  \Doi{DOI: 10.1016/j.aop.2004.01.008}{\bibinfo {journal} {Annals of Physics}}\
  }%
  \textbf{\bibinfo {volume} {312}},\ \bibinfo {pages} {354 -- 397} (\bibinfo
  {year} {2004}),\ ISSN \bibinfo {issn} {0003-4916},\
  \url{http://www.sciencedirect.com/science/article/B6WB1-4BSVNRB-1/2/30b4f75b%
eec12d3421907c2656059c94}\BibitemShut{NoStop}%
\bibitem{PhysRevA.68.014101}%
  \BibitemOpen
  \bibfield{author}{%
  \bibinfo {author} {\bibfnamefont{Merced}\ \bibnamefont{Montesinos}},\ }%
  \bibfield{title}{%
  \enquote{\bibinfo {title} {Heisenberg's quantization of dissipative
  systems},}\ }%
  \bibfield{journal}{%
  \Doi{10.1103/PhysRevA.68.014101}{\bibinfo {journal} {Phys. Rev. A}}\ }%
  \textbf{\bibinfo {volume} {68}},\ \bibinfo {pages} {014101} (\bibinfo {month}
  {Jul}\ \bibinfo {year} {2003})\BibitemShut{NoStop}%
\bibitem{chandrasekar:032701}%
  \BibitemOpen
  \bibfield{author}{%
  \bibinfo {author} {\bibfnamefont{V.~K.}\ \bibnamefont{Chandrasekar}},
  \bibinfo {author} {\bibfnamefont{M.}~\bibnamefont{Senthilvelan}},\ and\
  \bibinfo {author} {\bibfnamefont{M.}~\bibnamefont{Lakshmanan}},\ }%
  \bibfield{title}{%
  \enquote{\bibinfo {title} {On the lagrangian and hamiltonian description of
  the damped linear harmonic oscillator},}\ }%
  \bibfield{journal}{%
  \Doi{10.1063/1.2711375}{\bibinfo {journal} {Journal of Mathematical
  Physics}}\ }%
  \textbf{\bibinfo {volume} {48}},\ \bibinfo {eid} {032701} (\bibinfo {year}
  {2007}),\
  \url{http://link.aip.org/link/?JMP/48/032701/1}\BibitemShut{NoStop}%
\bibitem{Dito2006309}%
  \BibitemOpen
  \bibfield{author}{%
  \bibinfo {author} {\bibfnamefont{Giuseppe}\ \bibnamefont{Dito}}\ and\
  \bibinfo {author} {\bibfnamefont{Francisco~J.}\ \bibnamefont{Turrubiates}},\
  }%
  \bibfield{title}{%
  \enquote{\bibinfo {title} {The damped harmonic oscillator in deformation
  quantization},}\ }%
  \bibfield{journal}{%
  \Doi{DOI: 10.1016/j.physleta.2005.12.013}{\bibinfo {journal} {Physics Letters
  A}}\ }%
  \textbf{\bibinfo {volume} {352}},\ \bibinfo {pages} {309 -- 316} (\bibinfo
  {year} {2006}),\ ISSN \bibinfo {issn} {0375-9601},\
  \url{http://www.sciencedirect.com/science/article/B6TVM-4HSRP3H-2/2/4c37bef6%
57440f30fb2c96bae94c1804}\BibitemShut{NoStop}%
\bibitem{Kochan2010219}%
  \BibitemOpen
  \bibfield{author}{%
  \bibinfo {author} {\bibfnamefont{Denis}\ \bibnamefont{Kochan}},\ }%
  \bibfield{title}{%
  \enquote{\bibinfo {title} {How to quantize forces (?): An academic essay on
  how the strings could enter classical mechanics},}\ }%
  \bibfield{journal}{%
  \Doi{DOI: 10.1016/j.geomphys.2009.09.014}{\bibinfo {journal} {Journal of
  Geometry and Physics}}\ }%
  \textbf{\bibinfo {volume} {60}},\ \bibinfo {pages} {219 -- 229} (\bibinfo
  {year} {2010}),\ ISSN \bibinfo {issn} {0393-0440},\
  \url{http://www.sciencedirect.com/science/article/B6TJ8-4XDCHN8-1/2/ed88d72d%
5e7e026557c477b9416a744e}\BibitemShut{NoStop}%
\bibitem{PhysRevA.81.022112}%
  \BibitemOpen
  \bibfield{author}{%
  \bibinfo {author} {\bibfnamefont{Denis}\ \bibnamefont{Kochan}},\ }%
  \bibfield{title}{%
  \enquote{\bibinfo {title} {Functional integral for non-lagrangian systems},}\
  }%
  \bibfield{journal}{%
  \Doi{10.1103/PhysRevA.81.022112}{\bibinfo {journal} {Phys. Rev. A}}\ }%
  \textbf{\bibinfo {volume} {81}},\ \bibinfo {pages} {022112} (\bibinfo {month}
  {Feb}\ \bibinfo {year} {2010})\BibitemShut{NoStop}%
\bibitem{2009IJMPA..24.5319K}%
  \BibitemOpen
  \bibfield{author}{%
  \bibinfo {author} {\bibfnamefont{D.}~\bibnamefont{{Kochan}}},\ }%
  \bibfield{title}{%
  \enquote{\bibinfo {title} {{Quantization of Non-Lagrangian Systems}},}\ }%
  \bibfield{journal}{%
  \Doi{10.1142/S0217751X0904748X}{\bibinfo {journal} {International Journal of
  Modern Physics A}}\ }%
  \textbf{\bibinfo {volume} {24}},\ \bibinfo {pages} {5319--5340} (\bibinfo
  {year} {2009})\BibitemShut{NoStop}%
\bibitem{2007hep.th....3073K}%
  \BibitemOpen
  \bibfield{author}{%
  \bibinfo {author} {\bibfnamefont{D.}~\bibnamefont{{Kochan}}},\ }%
  \bibfield{title}{%
  \enquote{\bibinfo {title} {{Direct quantization of equations of motion: from
  classical dynamics to transition amplitudes via strings}},}\ }%
  \bibfield{journal}{%
  \bibinfo {journal} {ArXiv High Energy Physics - Theory e-prints}}%
   (\bibinfo {month} {Mar.}\ \bibinfo {year} {2007})\BibitemShut{NoStop}%
\bibitem{2009arXiv0906.3062L}%
  \BibitemOpen
  \bibfield{author}{%
  \bibinfo {author} {\bibfnamefont{T.}~\bibnamefont{{Luo}}}\ and\ \bibinfo
  {author} {\bibfnamefont{Y.}~\bibnamefont{{Guo}}},\ }%
  \bibfield{title}{%
  \enquote{\bibinfo {title} {{Infinite-Dimensional Hamiltonian Description of
  Dissipative Mechanical Systems}},}\ }%
  \bibfield{journal}{%
  \bibinfo {journal} {ArXiv e-prints}}%
   (\bibinfo {month} {Jun.}\ \bibinfo {year} {2009}),\
  \Eprint{http://arxiv.org/abs/0906.3062}{arXiv:0906.3062
  [math-ph]}\BibitemShut{NoStop}%
\bibitem{RickSalmon1988}%
  \BibitemOpen
  \bibfield{author}{%
  \bibinfo {author} {\bibfnamefont{Rick}\ \bibnamefont{Salmon}},\ }%
  \bibfield{title}{%
  \enquote{\bibinfo {title} {Hamiltonian fluid mechanics},}\ }%
  \bibfield{journal}{%
  \bibinfo {journal} {Ann. Rev. Fluid Mechanics}\ }%
  \textbf{\bibinfo {volume} {20}},\ \bibinfo {pages} {225--256} (\bibinfo
  {year} {1988})\BibitemShut{NoStop}%
\bibitem{RevModPhys.70.467}%
  \BibitemOpen
  \bibfield{author}{%
  \bibinfo {author} {\bibfnamefont{P.~J.}\ \bibnamefont{Morrison}},\ }%
  \bibfield{title}{%
  \enquote{\bibinfo {title} {Hamiltonian description of the ideal fluid},}\ }%
  \bibfield{journal}{%
  \Doi{10.1103/RevModPhys.70.467}{\bibinfo {journal} {Rev. Mod. Phys.}}\ }%
  \textbf{\bibinfo {volume} {70}},\ \bibinfo {pages} {467--521} (\bibinfo
  {month} {Apr}\ \bibinfo {year} {1998})\BibitemShut{NoStop}%
\bibitem{Morrison1980}%
  \BibitemOpen
  \bibfield{author}{%
  \bibinfo {author} {\bibfnamefont{P.~J.}\ \bibnamefont{Morrison}},\ }%
  \bibfield{title}{%
  \enquote{\bibinfo {title} {The maxwell-vlasov equations as a continuous
  hamiltonian system.}.}\ }%
  \bibfield{journal}{%
  \bibinfo {journal} {Phys. Lett. A}\ }%
  \textbf{\bibinfo {volume} {80}},\ \bibinfo {pages} {383--386} (\bibinfo
  {month} {Apr}\ \bibinfo {year} {1980})\BibitemShut{NoStop}%
\bibitem{Feynman1965}%
  \BibitemOpen
  \bibfield{author}{%
  \bibinfo {author} {\bibfnamefont{R.~P.}\ \bibnamefont{Feynman}}\ and\
  \bibinfo {author} {\bibfnamefont{A.~R.}\ \bibnamefont{Hibbs}},\ }%
  \emph{\bibinfo {title} {Quantum Mechanics and Path Integrals}}\ (\bibinfo
  {publisher} {McGraw-Hill},\ \bibinfo {year} {1965})\BibitemShut{NoStop}%
\bibitem{Feynman_thesis}%
  \BibitemOpen
  \bibfield{author}{%
  \bibinfo {author} {\bibfnamefont{Richard~Phillips}\ \bibnamefont{Feynman}},\
  }%
  \emph{\bibinfo {title} {A NEW APPROACH TO QUANTUM THEORY}},\ Ph.D. thesis,\
  \bibinfo {school} {Princeton University} (\bibinfo {year}
  {1942})\BibitemShut{NoStop}%
\bibitem{Das2005}%
  \BibitemOpen
  \bibfield{author}{%
  \bibinfo {author} {\bibfnamefont{U}~\bibnamefont{Das}}, \bibinfo {author}
  {\bibfnamefont{S}~\bibnamefont{Ghosh}}, \bibinfo {author}
  {\bibfnamefont{P}~\bibnamefont{Sarkar}},\ and\ \bibinfo {author}
  {\bibfnamefont{B}~\bibnamefont{Talukdar}},\ }%
  \bibfield{title}{%
  \enquote{\bibinfo {title} {Quantization of dissipative systems with friction
  linear in velocity},}\ }%
  \bibfield{journal}{%
  \bibinfo {journal} {Physica Scripta}\ }%
  \textbf{\bibinfo {volume} {71}},\ \bibinfo {pages} {235} (\bibinfo {year}
  {2005}),\
  \url{http://stacks.iop.org/1402-4896/71/i=3/a=002}\BibitemShut{NoStop}%
\end{thebibliography}
\end{document}